\documentclass[11pt,a4paper]{article}

\newcommand{\Blackhat}{{\sc BlackHat}}

\usepackage{jheppub}
\usepackage{amsmath}
\usepackage{amsfonts}
\usepackage{amssymb}
\usepackage{tikz}
\usetikzlibrary{arrows,shapes}

\usepackage[compat=1.1.0]{tikz-feynman}

\usepackage{amssymb}
\usepackage{color}
\begin{document}

\preprint{IPPP/19/28}

\title{Extracting analytical one-loop amplitudes from numerical evaluations}

\author{Giuseppe De Laurentis,}
\emailAdd{giuseppe.de-laurentis@durham.ac.uk}

\author{Daniel Ma\^{\i}tre}
\emailAdd{daniel.maitre@durham.ac.uk}

\affiliation{Institute for Particle Physics Phenomenology, Durham University}

\date{\today}

\abstract{
In this article we present a method to generate analytic expressions for the integral coefficients of loop amplitudes using numerical evaluations only. We use high-precision arithmetics to explore the singularity structure of the coefficients and decompose them into parts of manageable complexity. To illustrate the usability of our method we provide analytical expressions for all helicity configurations of the colour-ordered six-point gluon amplitudes at one loop with a gluon in the loop. 
}


\maketitle

\section{Introduction}

Numerical methods have been used in the calculation of scattering amplitudes when analytical calculations became intractable. A common bottleneck that makes  computations unfeasible is in intermediary steps rather than in the complexity of the final answer. A notable example of this is the Parke-Taylor formula for MHV amplitudes~\cite{ParkTaylor} which is significantly simpler than the intermediate expressions 
needed to calculate it. In this article we propose a method to recover the analytical form of expressions when only a numerical program is available for their evaluation. For example, this is often the case for high-multiplicity one-loop amplitudes.

Analytical expressions are often preferable to numerical solutions when they are to be used in extreme phase space configurations, such as for the integration in soft or collinear regions of the phase space. Analytical expressions can be expanded analytically in the relevant limit to provide numerically stable results. Furthermore, compact analytical expressions evaluate faster and with a smaller memory footprint than numerical procedures and can be more amenable to parallelisation.

The use of numerical samples to reconstruct analytical expressions is beginning to find direct applications to scattering amplitude calculations as a means of taming the complexity of the problem. In particular, computations over finite fields are used to perform integral reduction~\cite{vonManteuffel:2014ixa,Maierhoefer:2017hyi,Smirnov:2019qkx,Klappert:2019emp}, to reconstruct polynomials in kinematic variables in the calculation of two-loop QCD~\cite{Peraro:2016wsq,Zeng:2017ipr,Abreu:2017xsl,Badger:2017jhb,Abreu:2017hqn,Liu:2018dmc,Abreu:2018gii,Abreu:2018jgq,Badger:2018enw,Badger:2018gip,Abreu:2018zmy,Abreu:2018aqd,Abreu:2019odu} and $\mathcal{N}=8$ ~\cite{Chicherin:2019xeg,Abreu:2019rpt} amplitudes, as well as in higher loop calculations~\cite{Henn:2019rmi,vonManteuffel:2016xki}. The method described in this article differs from the above in that it uses large-precision floating-point arithmetics, rather than exact integer arithmetics modulo a prime number. This approach offers an easy interface to existing code as many programs already make use of high precision floating-point arithmetics to deal with numerical instabilities. As we will see later, using large scale differences allows us to restrict the calculation to specific parts of the answer rather than solving for the full answer at once. This targeted approach decreases the size of the fitting problem significantly.

To illustrate the usefulness of our method, we present analytical expressions for color-ordered six-gluon one-loop amplitudes with a gluon in the loop for all helicity configurations. These amplitudes were already calculated in the literature~\cite{Mahlon:1993si,Bern:1993qk,Bern:1994zx,Bern:1994cg,Bidder:2004tx,Bedford:2004nh,Britto:2005ha,Bern:2005cq,Bern:2005hh,Britto:2006sj,Berger:2006ci,Berger:2006vq,Xiao:2006vt} and summarised in Ref.~\cite{Dunbar:2009uk}, but to the best of our knowledge they were never presented in a single place using a single framework. Furthermore, the expressions we provide are explicitly rational (no square roots of spinors) and gauge invariant (no arbitrary reference momenta).

This article is organised as follows. Section~\ref{sec:method} presents the elements of our method. Section~\ref{sec:strategies} lays out the different ways of combining them into reconstruction strategies. Section~\ref{sec:results} describes the analytical determination of the one-loop scalar integral coefficients for the six-gluon one-loop amplitude with a gluon in the loop. Section~\ref{sec:conclusion} presents our conclusions.

\section{Method}\label{sec:method}
In this article we consider reconstructing an analytical expression for a rational quantity $\mathcal{E}$ for which we can calculate numerical values for arbitrary kinematics with arbitrary accuracy. 

We assume that the input to the program are the complex valued two-dimensional spinors $\lambda_i$ and $\tilde \lambda_i$ related to the external momenta $p_i$, which we assume to be all massless. Using complex momenta we can treat $\lambda_i$ and $\tilde \lambda_i$ as separate independent variables. This allows us to explore the singularity structure of our expression in a more controlled fashion.  
  
The method is based on the iteration of the following steps:
\begin{enumerate}
\item evaluate $\mathcal{E}$ in singular limits to obtain the list of all factors in the least common denominator (LCD) and their exponents;
\item consider $\mathcal{E}$ in doubly singular limits to expose the dependency structure of the poles;
\item select a pole from the LCD and identify the set of necessary other factors needed in the denominator to fit its residue;
\item subtract the term thus obtained from $\mathcal{E}$ and reiterate from step 1.
\end{enumerate}
At every iteration, at least a pole is either removed or its power reduced. We repeat the process until the expression is fully reconstructed.

The following sections explain the elements of the method in more details.

\subsection{Singular limits and least common denominator}
A rational expression $\mathcal{E}$ can be expressed over a single denominator
\begin{equation}\label{eq:lcd}
\displaystyle \mathcal{E}=\frac{\mathcal{N}}{{\mathcal{D}}_{\mathcal{LCD}}},
\end{equation}
where ${\mathcal{D}}_{\mathcal{LCD}}=\prod r^{n_i}_i$ denotes the least common denominator and $r_i$ the real poles of $\mathcal{E}$. $\mathcal{D}_{\mathcal{LCD}}$ is unique and can be obtained directly from numerical evaluations, given a sufficiently complete set $\{r_{i}\}$ of possible denominator factors, by numerically probing $\mathcal{E}$ in limits where one of these factors vanishes. We can construct a set of independent limits
\begin{equation}\label{eq:single_collinear_limit}
  r_{i}\rightarrow \epsilon \ll 1,\quad r_{j\neq i}=\mathcal{O}(1) \;, \end{equation}
to determine the powers of the factors $r_i$ by considering the scaling of $\mathcal{E}$ in the limits above. 

The same procedure also exposes overall factors in the numerator. In this case, $\mathcal{E}$ vanishes when the limit is taken rather than exhibiting the diverging behaviour of denominator factors. To test for overall factors we can use a broader set of structures $f_i$ that are not necessarily possible poles but satisfy the uniqueness of the limit in Eq.~(\ref{eq:single_collinear_limit}) in order to reliably ascribe the vanishing of the expression to a single overall factor. 

If $\mathcal{E}$ is a rational coefficient of scalar loop integrals in a one-loop scattering amplitude, the $f_i$'s are contractions of spinors and the limits of Eq.~(\ref{eq:single_collinear_limit}) correspond to a generalisation of collinear limits for complexified phase space. In the simplest cases, the procedure described in this section yields the full expression $\mathcal{E}$ up to a numerical prefactor, which can be obtained by performing a simple division. For example, this happens with all box coefficients in the six-gluon amplitudes.

\subsection{Doubly singular limits and partial fractions}
With large numbers of factors in the LCD, fitting the single numerator $\mathcal{N}$ becomes quickly intractable. Fortunately, we can often represent the expression as a sum of terms whose denominators have fewer factors or factors with a lower degree. In fact, such representations are often more compact and better represent the singularity structure of the expression.

Let us reconsider the expression $\mathcal{E}$ of Eq.~(\ref{eq:lcd}) now written as a sum of terms:
\begin{equation}\label{eq:edecomposition}
\displaystyle \mathcal{E}=\sum_{i} \frac{\mathcal{N}_i}{\mathcal{R}_{i}\mathcal{S}_{i}}.
\end{equation}
In the above, $\mathcal{R}_{i}$ are products of subsets of the factors in $\mathcal{D}_{\mathcal{LCD}}$, and $\mathcal{S}_{i}$ contain denominator factors that are not in the LCD, i.e.\ they cancel in the sum. The latter are known as spurious poles and arise naturally when using partial fractions to separate individual factors in the LCD. $\mathcal{N}_i$ are some numerator structures typically simpler than $\mathcal{N}$. Since the decomposition of Eq.~(\ref{eq:edecomposition}) is not unique, it can be used to optimise the compactness of the expression representation.

Insights on the possible structures of Eq.~(\ref{eq:edecomposition}) can again be obtained from particular regions of phase space. In analogy with the singular limits described in Eq.~(\ref{eq:single_collinear_limit}), we define doubly singular limits as:
\begin{equation}\label{eq:double_collinear_limit} 
  f_{i}\rightarrow \epsilon\ll 1, \quad f_{j}\rightarrow \epsilon \ll 1 
\end{equation}
and observe the behaviour of the expression $\mathcal{E} \sim \epsilon^{-n_{ij}}$ in this limit. In this case, it is not possible to guarantee that $f_{k\neq i,j}=\mathcal{O}(1)$. For example, if we have:
\begin{equation}
\langle 1|2 \rangle \rightarrow \epsilon \quad \mbox{and} \quad \langle 2|3 \rangle \rightarrow \epsilon 
\end{equation}
we must also have:
\begin{equation}
\langle 1|3 \rangle \sim \epsilon \,,\; s_{123} \sim \epsilon\,,\; ...
\end{equation}
We call the double limit Eq.~(\ref{eq:double_collinear_limit}) \emph{clean} if no factor $f_{k\neq i,j}$ other than $f_i$ and $f_j$ vanishes in this limit. 

The singular limit of Eq.~(\ref{eq:double_collinear_limit}) is symmetric: $f_i$ and $f_j$ are both set to the same small $\epsilon$. However, in some cases it is useful to study asymmetric limits as well: $f_i \rightarrow \epsilon_i$, $f_j \rightarrow \epsilon_j$, $\epsilon_i \neq \epsilon_j$. This is especially important to lift degeneracies that arise with higher order poles.

The two most interesting sets of doubly singular limits are: a) for ($f_i$, $f_j$) both real poles ($r_i$, $r_j$); and b) for $f_i$ a real pole $r_i$, and $f_j \notin \mathcal{D}_{\mathcal{LCD}}$. Let us start with the former case and for the sake of simplicity an expression $\mathcal{E}$ which only involves simple poles, such as tree amplitudes. The reasoning for expressions involving higher order poles is similar. There are three distinct cases:
\begin{enumerate}
\item the limit Eq.~(\ref{eq:double_collinear_limit}) is clean
and $n_{ij}=1$: this implies that we can find a representation for $\mathcal{E}$ where $r_i$ and $r_j$ never appear in the same denominator. They can be split up without the need of a spurious pole;
\item the limit Eq.~(\ref{eq:double_collinear_limit}) is clean
%$f_{k\neq i,j}=\mathcal{O}(1)$ 
and $n_{ij}=2$: this implies that $r_i$ and $r_j$ must appear at least once in  the same denominator in the sum of Eq.~(\ref{eq:edecomposition}) and our set $\{f\}$ does not contain a spurious pole able to separate them;
\item the limit Eq.~(\ref{eq:double_collinear_limit}) is not clean, i.e.~there exist vanishing factors $v_k\sim\epsilon$ in the double limit:
  \begin{itemize}
  \item if $n_{ij}=1$ we cannot numerically distinguish the following situations:
    \begin{equation}
      \mathcal{E} \sim \frac{v_k}{r_ir_j}\;, \quad \mathcal{E} \sim \frac{1}{r_i}\;, \quad \mathcal{E} \sim \frac{1}{r_j};
    \end{equation}
  The implication is that in this case we cannot conclude from the doubly singular limit whether $r_i$ and $r_j$ have to be present at the same time in a denominator. 
  \item if $n_{ij}=2$ we cannot numerically distinguish the following situations:
    \begin{equation}
      \mathcal{E} \sim \frac{1}{r_ir_j}\;, \quad \mathcal{E} \sim \frac{1}{r_i v_k}\;, \quad \mathcal{E} \sim\frac{1}{r_jv_k};
    \end{equation}
  and linear combinations of these scenarios are also possible. 
  \end{itemize}
\end{enumerate}
The consequence of not being able to discriminate between these scenarios is that multiple possible ansatze for the denominator structure are possible and, as mentioned before, there is often no obvious optimal solution.

Let us consider the latter expression when $v_k$ does not appear in $\mathcal{D}_{\mathcal{LCD}}$. There might be several distinct $v_k$'s and among those one may recognise some as possible spurious poles $s_k$, which now have a clear physical interpretation in that they preserve the correct doubly singular behaviour of each term when we separate the $r_i$ and $r_j$ poles in separate denominators. 

For instance, let us consider the structure of the pair of poles $r_i$, $r_j$ = $\langle 12\rangle$, $[34]$ when $n_{ij}=2$. Among the $f_k$'s vanishing in this double singular limit, the relevant spurious pole $s_k$ is $\langle 1|2+3|4]$, and the partial fraction identity reads:
\begin{equation}
  \frac{1}{\langle 12\rangle [34]} = \frac{\langle 13\rangle}{\langle 12\rangle\langle 1|2+3|4]} + \frac{[24]}{[34] \langle 1|2+3|4]}
\end{equation}
We can see how the spurious pole prevents $\langle 1| 2\rangle$ and $[3|4]$ from appearing in the same denominator in the well known representation of $A_{tree}^{\scriptsize{+-+-+-}}$:
\begin{eqnarray}\label{eq:ApmpmpmOne}
A_{tree}(1^+,2^-,3^+,4^-,5^+,6^-)&=&
\phantom{+}\frac{1i\langle2|1+3|5]^4}{\langle12\rangle\langle23\rangle[45][56]\langle1|2+3|4]\langle3|1+2|6]s_{123}}\nonumber\\
&&+\frac{1i\langle6|2+4|3]^4}{\langle16\rangle[23][34]\langle56\rangle\langle1|2+3|4]\langle5|1+6|2]s_{234}}
\nonumber\\&&+
\frac{-1i\langle4|3+5|1]^4}{[12][16]\langle34\rangle\langle45\rangle\langle3|1+2|6]\langle5|1+6|2]s_{345}}.
\end{eqnarray}
We can also see that $\langle 3|1+2|6]$ separates $\langle2|3\rangle$ and $[1|6]$, and $\langle 5|1+6|2]$ separates $\langle5|6\rangle$ and $[1|2]$. However, another representation is also possible where $\langle 12\rangle$ and $[34]$ appear in the same denominator and different spurious poles separate other pairs of poles:
\begin{eqnarray}\label{eq:ApmpmpmTwo}
A_{tree}(1^+,2^-,3^+,4^-,5^+,6^-)&=&
\phantom{+}\frac{1i[13]^4\langle46\rangle^4}{[12][23]\langle45\rangle\langle56\rangle\langle4|2+3|1]\langle6|1+2|3]s_{123}}\nonumber\\
&&+
\frac{1i[15]^4\langle24\rangle^4}{[16]\langle23\rangle\langle34\rangle[56]\langle2|1+6|5]\langle4|2+3|1]s_{234}}\nonumber\\
&&+
\frac{-1i\langle26\rangle^4[35]^4}{\langle12\rangle\langle16\rangle[34][45]\langle2|1+6|5]\langle6|1+2|3]s_{345}}.
\end{eqnarray}

A good choice of spurious pole $s_k$ to introduce can be identified from the list of vanishing $v_k$'s by looking at intersections between different $v_k$ sets: for instance, the $\{v\}$ sets from ($\langle 12\rangle$, $[34]$) and ($\langle 16\rangle$, $[45]$) share $\langle 1|2+3|4]$ only.

The other interesting set of doubly singular limits is for $f_i$ a real pole $r_i$ while $f_j \notin \mathcal{D}_{\mathcal{LCD}}$. Whenever in such a limit $\mathcal{E}$ diverges less drastically than $n_i$, that is $n_{ij} < n_i$ where $n_i$ is the order of the single pole $r_i$, it means that in this doubly singular limit $f_j$ is a factor in the numerator of the term containing $r_i$ in the denominator. As explained above, more or less information can be accessed this way depending on the degeneracy of the particular phase space point, i.e.\ how many $v_k\neq f_i,f_j$ vanish in this limit.

Using these observations we can write an ansatz for the expression in the form of Eq.~(\ref{eq:edecomposition}) where the denominators $\mathcal{D}_i=\mathcal{R}_i\mathcal{S}_i$ are free from combinations of factors that would lead to a worse scaling than observed in the doubly singular limits. Alternatively, we can attempt to express $\mathcal{E}$ as a sum of terms apparently violating the doubly singular scalings, but free from spurious poles.

\subsection{Numerator ansatz and coefficients reconstruction}

In this section we discuss how to reconstruct a numerator whenever singular limits do not provide all of the required information. We start by building an ansatz out of products of spinor products $\left<i|j\right>$ and $\left[i|j\right]$ and, if necessary, other linearly independent expressions, such as square roots of Gram determinants. The coefficients of the terms in the ansatz are determined by solving a system of linear equations. The number of spinor products in each term of the ansatz can be determined by numerically inspecting the behaviour of the expression under uniform scaling of all momenta:
\begin{equation}
p_i\rightarrow \lambda p_i\;,
\qquad 
\left<i|j\right>\rightarrow \lambda \left<i|j\right> 
\;,
\qquad 
\left[i|j\right]\rightarrow \lambda \left[i|j\right],
\qquad
\forall \; i, j \in \{1, n\}.
\end{equation}

We will refer to the power of $\lambda$ in an expression as its \emph{mass dimension}. Alternatively, we may think of the mass dimension as the degree of the polynomial in the angle and square brackets, since they have mass dimension $1$. We can further limit the size of the ansatz by looking at its \emph{phase weights}. The phase weight with respect to momentum $p_i$ is defined by the scaling of the expression under a little group transformation, i.e.~a change in  $\lambda_i$ and $\tilde \lambda_i$ which leaves $p_i$ unchanged:
\begin{equation}
p_i\rightarrow p_i\;,\qquad |i\rangle\rightarrow \phi \,|i\rangle\;,\qquad [i|\rightarrow \phi^{-1} \, [i|.
\end{equation}
The phase weight for momentum $i$ of the expression $\mathcal{E}$ is $n$ if it scales as $\phi^n$. The phase-weight of the numerator ansatz combined with that of the denominator has to match the phase-weight of the expression. 

An ansatz built from all products of spinor products with the right mass dimension and phase weights is sufficient but not minimal, due to momentum conservation and Schouten identities. To ensure uniqueness of the numerator representation, we need to remove redundant elements from the ansatz by either using analytical rules or numerical Gaussian elimination. This operation only needs to be performed once per mass dimension and phase weights combination.

For our application we use a numerical Gaussian elimination implemented in the following way: for a candidate ansatz $\mathcal{A}$ with $N$ elements $a_{j=1,...,N}$ we generate $N$ distinct phase space points $P_i$ and build a $N\times N$ matrix $M$. The set of linear identities $\{v\}$ relating ansatz elements lives in the kernel of $M$, that is for each such identity $v$ for which $\sum_{j} a_j v_j = 0 $
we have: 
\begin{equation}\label{eq:independence}
M_{ij}v_{j}=0\,\quad \mbox{with}\quad M_{ij} = a_{j}(P_i)\;.
\end{equation}
Here we are not interested in the identities but merely wish to remove redundant ansatz elements that can be expressed in terms of other elements. Row-reducing $M$ brings it in upper-triangular form and the existence of identities will manifest itself as the appearance of zeros in the diagonal of the transformed matrix. As the algorithm progresses, we remove each ansatz element that leads to a vanishing diagonal element in the transformed matrix and remove the corresponding column of $M$. At the end of the row-reduction procedure we are left with $N'\leq N$ elements in the ansatz. These $N'$ elements are linearly independent.

Given a minimal ansatz, we can solve for the coefficients vector $c$ of each term in the numerator by solving the equation
\begin{equation}\label{eq:inversion}
\sum_j\tilde M_{ij}\, c_{j} =  \tilde a_{j}(P_i)\, c_j = \mathcal{E}(P_i)\,,\quad \Rightarrow \quad c_j=\sum_i\tilde M^{-1}_{ji}\mathcal{E}(P_i)
\end{equation}
where $\tilde M$ is the matrix of the $N'$ independent numerator ansatz elements $\tilde a$ that we constructed above through Gaussian elimination, divided by the corresponding denominator. $\mathcal{E}(P_i)$ is the vector of the expression $\mathcal{E}$ evaluated at the first $N'$ phase space configurations. In all cases we considered the coefficients in $c$ are expected to be rational numbers. The analytical, infinite precision values can be recovered from the numerical estimates obtained through the inversion in Eq.~(\ref{eq:inversion}) with procedures such as that of continued fractions. One can easily check the validity of the expression obtained by testing a further distinct phase space point. Note also that the inverse $\tilde M^{-1}$ is not explicitly calculated: large matrix inverses constructed numerically are susceptible to instabilities. Instead, Eq.~(\ref{eq:inversion}) is solved through the same row-reduction procedure used for Eq.~(\ref{eq:independence}).

\section{Reconstruction strategies}\label{sec:strategies}
Depending on the complexity of the expression to reconstruct, we can apply different strategies:
\begin{itemize}
\item[a)] full reconstruction,
\item[b)] full reconstruction with separated denominators,
\item[c)] iterated reconstruction by sequentially removing poles.
\end{itemize}

\subsection{Full reconstruction}\label{sec:fullreconstruction}
Strategy a) is the simplest and does not require doubly singular limits to be probed. Unfortunately, trying to solve for the numerator $\mathcal{N}$ of the least common denominator $\mathcal{D}_{\mathcal{LCD}}$ is in general intractable. The mass dimension of $\mathcal{N}$ can easily exceed 12, with the worst of the six-point amplitudes coefficients being above 100. Table~\ref{tab:indepterms} shows the size of the minimal ansatz $\{\tilde a\}$ as a function of mass dimension at six-point for constant phase weights $[0, 0, 0, 0, 0, 0]$.

\begin{table}
\begin{center}
\begin{tabular}{r|c|c|c|c|c|c}
mass dimension & 2 & 4 & 6 & 8 & 10 & 12\\ \hline
independent terms & 9 & 50 & 205 & 675 & 1886 & 4644\\
\end{tabular}
\end{center}
\caption{\label{tab:indepterms}Number of independent terms in an ansatz for six-momentum configurations with all zero phase weights as a function of the mass dimension.}
\end{table}

\subsection{Full reconstruction with separated denominator}
For strategy b) we use the information from doubly singular limits to postulate possible sets of denominators to write $\mathcal{E}$ as a sum of terms with simpler denominators ${\mathcal D}_i$, possibly containing spurious poles. \begin{equation}\label{eq:sum-of-terms}
\mathcal{E}=\sum_i\frac{\mathcal{N}_i}{\mathcal{D}_i}
\end{equation}  
There are different ways of choosing the denominators ${\mathcal D}_i$ depending on the number of terms in the sum and which spurious poles are chosen. 

We apply the technique described above to construct an ansatz for each numerator 
\begin{equation}
{\mathcal N}_i = \sum_{j=1}^{N_i} c_{i,j}\frac{a_{i,j}}{\mathcal{D}_i}
\end{equation}
where $N_i$ is the number of elements in the ansatz for the $i^{th}$ numerator $\mathcal{N}_i$. Generally the combined size of these ansatze is much less than that for the single numerator $\mathcal{N}$.

While each numerator ansatz is constructed with independent elements, the sum over the terms can still contain redundant terms. For example, if we have
\begin{equation}
\mathcal{E} = \frac{\mathcal{N}_A}{AB}+\frac{\mathcal{N}_B}{AC}\;,
\end{equation}
a term proportional to $BC$ in the numerator $\mathcal{N}_A$ can be moved to $\mathcal{N}_B$ and viceversa.

This redundancy can be removed with the same technique described above. In analogy to Eq.~(\ref{eq:independence}), we construct $N=\sum N_i$ distinct momentum configurations $P_l$ and a matrix $M$:
\begin{equation}
M_{l,k} = A_{k}(P_l) \;,\qquad A_k(P_l) =\frac{a_{i,j}(P_l)}{\mathcal{D}_{i}(P_l)}\;,\qquad k=i + \sum_{\tilde i < i}N_{\tilde i}
\end{equation}  
where $a_{i,j}$ is the $j^{th}$ element of the ansatz of the $i^{th}$ numerator, and $k$ enumerates through the elements of the combined ansatze of the terms in Eq.~(\ref{eq:sum-of-terms}). Redundant elements are then removed with the row-reduction procedure. Once the ansatz is minimal, we can solve for the coefficients $c_{i,j}$ by inverting a numerical system of equation.

Even by separating the LCD into smaller denominators, the resulting system can get too large to be solved in a reasonable time. In the following sections we discuss two methods to resolve this issue through the use of singular limits and symmetries.

\subsection{Iterated reconstruction by sequentially removing poles}\label{sec:iteratedreconstruction}

For expressions for which strategies a) and b) are intractable, we use the full method presented in Section~\ref{sec:method}. The aim is to isolate the contribution of the highest order of a specific pole $r$ in the expression $\mathcal{E}$. To achieve this we identify the term $i_{r}$ in Eq.~(\ref{eq:edecomposition}) with the highest power $k$ of the pole $r$, that is, we think of $\mathcal{E}$ in the form:
\begin{equation}
{\mathcal E} = \frac{{\mathcal N}_{i_r}}{r^k \bar{\mathcal D}_{i_r}}+\sum\limits_{i\neq i_r} \frac{{\mathcal N}_i}{r^{k_i} \bar{\mathcal D}_i} \,,
\end{equation} 
where the powers of $r$ in the denominators in the sum are lower than in the first term, i.e.\ $k_i< k$, and $\bar {\mathcal D}$ are the denominators with any power of the pole $r$ factored out, i.e.\ ${\mathcal D}_i = r^{k_i}\bar {\mathcal D}_i$.

We can fit the numerator $\mathcal{N}_{i_r}$ in isolation if we generate the phase space points for the Gaussian elimination in the specific singular limit $r \rightarrow \epsilon$, thus making the $i_r$ term dominant. Subtracting the term reconstructed in this limit from $\mathcal{E}$ results in an expression where the order of the pole $r$ is decreased by one. Repeating the same operation for the new maximum power of $r$ or for other factors reduces the mass dimension of the numerators until the remaining expression can be fitted without any particular limits using the strategies a) or b).

There is a large amount of freedom in choosing the order in which to remove the poles, which can lead to very different forms of the reconstructed analytical expression. This freedom can be exploited for different goals. On the one hand, we can iterate through different choices to select the most compact version, in order to obtain the quickest evaluation. On the other hand, we can produce expressions that are numerically stable in specific limits by either removing the poles corresponding to the selected singular behaviour first or by avoiding the introduction of certain spurious singularities. In doing so we can produce a family of expressions, each tailored to maximise execution speed or numerical stability in specific phase-space regions.  

\section{Six-gluon Results}\label{sec:results}

To illustrate our method we obtain analytical expressions for the scalar integral coefficients of the one-loop six-gluon amplitudes with a gluon in the loop. These amplitudes can be written in terms of scalar bubble, triangle and box integrals and a rational term as:
\begin{equation}
\mathcal{A}^{1-loop}_{6g} = \frac{\Gamma(1+\epsilon)\Gamma(1-\epsilon)^2}{(2\pi)^{2-\epsilon}\Gamma(1-2\epsilon)} \left( \sum\nolimits_i b_i I^4_i + \sum\nolimits_j c_j I^3_k + \sum\nolimits_k d_k I^2_k + R \right).
\end{equation}
In the above, $I^4_i$ are the scalar box integrals, $I^3_j$ the scalar triangle integrals and $I^2_k$ the scalar bubble integrals. The coefficients $b_i$, $c_j$, $d_k$ and $R$ are the rational functions of spinor products for which we applied our reconstruction method.

We emphasise that in extracting the analytical expressions for the coefficients we did not exploit any prior knowledge about the coefficients beyond the list of possible factors in the denominator. This list is a property of one-loop amplitudes with massless internal particles. Their powers and how they combine has been uncovered by the numerical exploration. More specifically, only knowledge about the general structure of these Lorentz invariants was used. We programmatically generate all strings of the form $s_{ijk}$, $\Delta_{ijk}$ (see Eq.~(\ref{eq:deltadefinition})), $\langle ij\rangle$, $[ij]$, $\langle i|j+k|l]$, and so forth.

We used the \Blackhat{} library \cite{Berger:2008sj} and its arbitrary precision implementation using the GNU Multi Precision library~\cite{GMP} to generate the numerical input for our method. These amplitudes were previously calculated numerically in Ref.~\cite{Ellis:2006ss} and combined to present the NLO four-jet cross section and distributions in Ref.~\cite{Bern:2011ep} and Ref.~\cite{Badger:2012pf}.

In the accompanying files we provide expressions readable by the {\sc S@M}~\cite{Maitre:2007jq} {\sc Mathematica}~package as well as human readable formulae for a representative set of helicity configurations. All other configurations can be obtained through symmetries.
%We have validated all of our analytical results by numerically comparing them to the output of \Blackhat{} on several independent phase space points to 300 significant digits.
We have validated all our analytical results by verifying their agreement with the output of \Blackhat{} to 300 significant digits on several independent phase space points (i.e.~phase space points that were not used in the determination of the coefficients of the ansatz). 
\subsection{Execution speed comparison}

For the reconstruction of the analytical expression for the integral coefficients we have treated each coefficient in isolation and did not use any knowledge about relationships between coefficients of related scalar integrals. This means that the resulting expressions could easily be re-written in a more compact way but we refrained from doing so as in the current form they are more illustrative of the type of output our method produces. This also provided us with additional validation methods for our results (for example, we checked that the sum of the bubble coefficients is proportional to the tree amplitude).

In order to assess the potential gain of using our analytical expression we implemented the analytical expressions in \Blackhat{}, which allows us to perform a comparison where the only difference is whether the numerical procedure or the analytical expressions are used. We observe significantly lower run times compared to the original numerical computation, with individual pieces receiving different speed-ups. The best speed improvement is by a factor of about $75$ for the split NMHV configuration, while the worst is a factor of $2$ for the alternating NMHV configuration. The remaining NMHV configuration is about $3$ times faster. In the latter two cases, the analytical formulae for the cut part of the amplitude led to slightly slower code. However, since the largest part of the calculation time in \Blackhat{} is spent on the rational part, which is significantly faster analytically, we still measure an overall speed-up for the complete amplitude. On the entire cross section the speed-up lies in between those of the various helicity configurations: it is a factor of about 4. Since the MHV and split NMHV configurations run much faster analytically than numerically, the bottlenecks for the entire cross section are the two harder NMHV configurations.

As pointed out earlier, the execution speed could be further improved if the expressions were simplified using some additional knowledge about the structure of the one-loop amplitude. Similarly, some post-processing of the reconstructed coefficients could be beneficial, but might misrepresent the method output. For instance, let us consider the following expression:
\begin{equation}\label{eq:compact}
\mathcal{E} = \frac{\langle1|2+3|5]\langle3|1+2|5]^3}{\langle13\rangle^4} + O(\langle13\rangle ^ 0).
\end{equation}
Regardless of how complicated the $O(\langle13\rangle ^ 0)$ part is, we can isolate the first term by considering phase space points in the $\langle13\rangle$ singular limit. However, since the expression above groups $\langle13\rangle$ sub-leading terms in $\langle1|2+3|5]$ and $\langle3|1+2|5]$, the reconstruction strategies presented in the previous section will yield a Laurent expansion in $\langle13\rangle$:

\begin{eqnarray}\label{eq:expanded}
  \mathcal{E}&=&-\frac{\langle12\rangle\langle23\rangle^3[25]^4}{\langle13\rangle^4} - \frac{\langle23\rangle^2[25]^3(3\langle12\rangle[15]+\langle23\rangle[35])}{\langle13\rangle^3}+\nonumber\\
  &&-\frac{[15]\langle23\rangle[25]^2(3\langle12\rangle[15]+3\langle23\rangle[35])}{\langle13\rangle^2}-\frac{[15]^2[25](\langle12\rangle[15]+3\langle23\rangle[35])}{\langle13\rangle}+\nonumber\\
  &&-[15]^3[35] + O(\langle13\rangle ^ 0)
\end{eqnarray}

The last spinor helicity term is actually itself $O(\langle13\rangle ^ 0)$, and thus it would need to be obtained independently of the $\langle13\rangle$ singular limit, but we reproduce it here for completeness. Clearly Eq.~(\ref{eq:compact}) would evaluate much faster than Eq.~(\ref{eq:expanded}).

% There are two ways in which we can attempt to recover compact resumed expression such as Eq.~(\ref{eq:compact}), but both involve some guess work. The first one is to feed back into the algorithm the relevant parts of the full expression, then factors like $\langle1|2+3|5]$ and $\langle3|1+2|5]$ will trivially appear as zeros of the sub-expression. However, it may not be easy to identify which part of the O($\langle13\rangle ^ 0$) terms to include. The second one is to consider more doubly singular limits. For instance, in the limit where both $\langle13\rangle$ and  $\langle1|2+3|5]$ vanish $\mathcal{E}$ goes to a constant. However, in such a limit also $\langle3|1+2|5]$ and $[25]$ vanish. This tell us that we need 4 powers of a combination of $\langle1|2+3|5]$, $\langle3|1+2|5]$ and $[25]$ in the term with the leading $\langle13\rangle$ singularity, but not which combination is the best one.

Lastly, in some cases we stop splitting the pole structure into smaller denominators when the full reconstruction of the numerator becomes feasible.
% (e.g.~in the alternating NMHV rational part, which is already several times faster than its numerical counterpart). However, this resulted in a couple of large terms which somewhat slow down the computation.
As future work, it might be interesting to try to further unravel the pole structure of such terms to potentially obtain more compact representations.

\subsection{Rationality of the one-loop coefficients}

It has already been shown in Ref.~\cite{BjerrumBohr:2007vu} that the coefficients of three-mass triangles in $\mathcal{N}=1$ super Yang-Mills can be written in a manifestly rational form at six-point. We observe that this holds also without any super-symmetry, and for bubble coefficients, and the rational part. To achieve this, we use information on the singularity structure of these quantities, which explains why square roots of Gram determinants seem to appear and why the same behaviour can be reproduced by rational spinor structures.

\begin{figure}
\begin{center}
  \begin{tikzpicture}
    \begin{feynman}
      \vertex [] (central) {};
      \vertex[blob, style={/tikz/minimum size=0.5cm}] [above = 1.15470cm of central] (top) {};
      \vertex [below right = 0.2cm and 0.5cm of top] (top_label_right) {$+$};
      \vertex [below left = 0.2cm and 0.5cm of top] (top_label_left) {$-$};
      \vertex [above left = 1.15470cm and 1cm of top] (top_left) {$1^+$};
      \vertex [above right = 1.15470cm and 1cm of top] (top_right) {$2^-$};
      \vertex[blob, style={/tikz/minimum size=0.5cm}] [below left = 0.57735cm and 1cm of central] (bottom_left) {};
      \vertex [above left = 0.5cm and 0.1cm of bottom_left] (bottom_left_label_above) {$+$};
      \vertex [below right = 0.3cm and 0.5cm of bottom_left] (bottom_left_label_below) {$-$};
      \vertex [below left = 1.15470cm and 1cm of bottom_left] (bottom_left_right) {$5^+$};
      \vertex [left = 1.5cm of bottom_left] (bottom_left_left) {$6^-$};
      \vertex[blob, style={/tikz/minimum size=0.5cm}] [below right = 0.57735cm and 1cm of central] (bottom_right) {};
      \vertex [above right = 0.5cm and 0.1cm of bottom_right] (bottom_right_label_above) {$-$};
      \vertex [below left = 0.3cm and 0.5cm of bottom_right] (bottom_right_label_below) {$+$};
      \vertex [below right = 1.15470cm and 1cm of bottom_right] (bottom_right_left) {$4^-$};
      \vertex [right = 1.5cm of bottom_right] (bottom_right_right) {$3^+$};
      \diagram* {
        (top) -- [gluon] (bottom_left),
        (top) -- [gluon] (top_left),
        (top) -- [gluon] (top_right),
        (bottom_left) -- [gluon] (bottom_right),
        (bottom_left) -- [gluon] (bottom_left_left),
        (bottom_left) -- [gluon] (bottom_left_right),
        (bottom_right) -- [gluon] (top),
        (bottom_right) -- [gluon] (bottom_right_left),
        (bottom_right) -- [gluon] (bottom_right_right),
      };
    \end{feynman}
  \end{tikzpicture}

\end{center}
\caption{Three-mass triangle.}\label{fig:NMHV}
\end{figure}
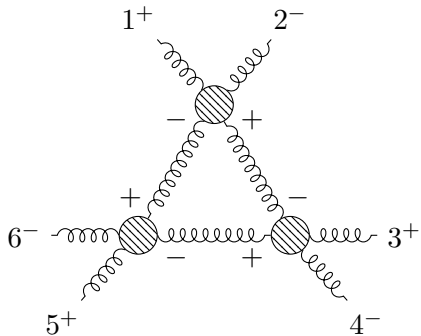

For concreteness, let us consider one of the two three-mass triangles in the alternating NMHV helicity configuration shown in Figure~\ref{fig:NMHV}.
The real poles of this function, as obtained from the singular limits of Eq.~(\ref{eq:single_collinear_limit}), are:
\begin{equation}\label{eq:pmpmpmthreemasstrianglerealpoles}
\langle12\rangle, [12], \langle34\rangle, [34], \langle56\rangle, [56], \langle1|3+4|2]^4, \langle3|1+2|4]^4, \langle5|1+2|6]^4, \Delta_{135}^3.
\end{equation}
Following convention from the literature, in the above list $\Delta_{135}$ is the Gram determinant related to this diagram:
\begin{equation}\label{eq:deltadefinition}
\Delta_{135} = (K_1\cdot K_2)^2-K_1^2K_2^2,
\end{equation}
where $K_1$ and $K_2$ are the sums of the momenta in any two corners of Figure~\ref{fig:NMHV}.

Square roots seem to appear when we study doubly singular limits, for instance:
\begin{equation}
\langle3|1+2|4] \rightarrow \epsilon,\;\Delta_{135} \rightarrow \epsilon,\; 
\quad \mbox{yields}\quad
-\frac{\log(\mathcal{E})}{\log(\epsilon)} \rightarrow 3.5,
\end{equation}
or:
\begin{equation}\label{eq:squareroot12}
\langle12\rangle \rightarrow \epsilon^3,\; \Delta_{135} \rightarrow \epsilon,\; \quad \mbox{yields}\quad
-\frac{\log(\mathcal{E})}{\log(\epsilon)} \rightarrow 2.5,
\end{equation}
and similarly for the other poles. Note how the asymmetric doubly singular limit in Eq.~(\ref{eq:squareroot12}) was necessary to lift the $1/\langle12\rangle$ residue above the $1/\Delta^3_{135}$ one. Although this scaling can be explained by an irrational factor of $\sqrt\Delta_{135}$, there is a more appealing solution: in any limit exhibiting half integer scaling, $\Delta_{135}$ behaves like the square of some rational quantity, and it is sufficient to introduce this quantity in the numerator instead of $\sqrt\Delta_{135}$. Several such spinor structures are possible. Here we list a few with their relation to $\Delta_{135}$:
\begin{equation}\label{eq:rootdeltafirst}
(\Omega_{351})^2 \equiv (2s_{12}s_{56}-(s_{12}+s_{56}-s_{34})s_{123})^2 = 4s_{123}^2\Delta_{135}-4s_{12}s_{56}\langle4|1+2|3]\langle3|1+2|4]
\end{equation}
\begin{equation}
(\Pi_{351})^2 \equiv (s_{123}-s_{124})^2 = 4\Delta_{135}-4\langle4|1+2|3]\langle3|1+2|4]
\end{equation}
\begin{equation}
(\langle1|3+4|1]+\langle2|3+4|2])^2 = 4\Delta_{135}+4s_{12}s_{34}
\end{equation}
\begin{equation}\label{eq:rootdeltalast}
-(s_{34}-s_{56})^2 = \Delta_{135}+ \langle12\rangle...
\end{equation}
The first two quantities might be familiar from the numerators of the expressions obtained in Ref.~\cite{BjerrumBohr:2007vu}. The order of the subscripts in those quantities is important because there are 3 distinct ones, one for each corner of the triangle, whereas $\Delta_{135}$ is invariant under a permutation of its subscripts. As a concrete example, the $1/\Delta^3_{135}$ term is almost fully constrained by doubly singular limits and can be expressed as:
\begin{equation}\label{eq:deltacubeterm}
\small\frac{5/128i\langle12\rangle[12]\langle34\rangle[34]\langle56\rangle[56]\langle2|3+4|1]\langle4|1+2|3]\langle6|1+2|5]\Pi_{135}\Pi_{351}\Pi_{513}}{\langle1|3+4|2]\langle3|1+2|4]\langle5|1+2|6]\Delta_{135}^3}.
\end{equation}

An issue with Eq.~(\ref{eq:deltacubeterm}) is that it introduces spurious singularities in doubly singular regions where a pair of the three poles $\langle1|3+4|2]$, $\langle3|1+2|4]$ and $\langle5|1+2|6]$ vanishes. This can be fixed by adding the following term:
\begin{equation}\label{eq:deltacubeterm_nospurious}
\small\frac{5/32i\langle12\rangle[12]\langle34\rangle[34]\langle56\rangle[56]\langle2|3+4|1]\langle4|1+2|3]\langle6|1+2|5](\Pi_{135}+\Pi_{351}+\Pi_{513})}{\langle1|3+4|2]\langle3|1+2|4]\langle5|1+2|6]\Delta_{135}^2}.
\end{equation}
Finally, for comparison, the same triple pole in the bubbles reads:
\begin{equation}
  \small\frac{5/256i\langle12\rangle[12]\langle2|3+4|1]\langle4|1+2|3]\langle6|1+2|5](s_{134}+s_{234})\Pi_{135}\Pi_{351}\Pi_{513}}{\langle1|3+4|2]\langle3|1+2|4]\langle5|1+2|6]\Delta_{135}^3},
\end{equation}
whereas in the rational part it enters at order $\Delta_{135}^2$ as:
\begin{equation}\label{eq:deltasquare_rational}
\small\frac{5/96i\langle2|3+4|1]\langle4|1+2|3]\langle6|1+2|5]\Pi_{135}\Pi_{351}\Pi_{513}}{\langle1|3+4|2]\langle3|1+2|4]\langle5|1+2|6]\Delta_{135}^2}.
\end{equation}

Expressions such as these in Eq.~(\ref{eq:deltacubeterm}-\ref{eq:deltasquare_rational}) exhibit a scaling consistent with square roots of $\Delta_{135}$ when considered in particular kinematical regions but are fully rational. More generally, the use of the spinor structures in Eq.~(\ref{eq:rootdeltafirst}-\ref{eq:rootdeltalast}) allowed us to obtain rational analytical representations for all the pieces of the one-loop six-gluon amplitudes. 
\subsection{Symmetries}
Symmetries of the coefficients also help in the analytical reconstruction. A coefficient may be invariant under a symmetry operation or two coefficients can be related by it. In the former case, the number of pole residues that have to be fitted is reduced: once a pole has been removed all other poles related to it by a symmetry can also be removed by a simple symmetrisation. In the latter case, the symmetries reduce the number of coefficients we have to consider: it is sufficient to consider independent topologies.

For pure gluon amplitudes symmetries are permutations of the external indices; they can be either cyclic of anti-cyclic, with anti-cyclic permutation involving a an overall factor of $(-1)^n$ from parity, where $n$ is the multiplicity of the phase space. Symmetries may involve a flip of all helicities, which corresponds to swapping the left and right Lorentz spinor representations. The latter operation is equivalent to complex conjugation in the case of real momenta.

Therefore, we can express a symmetry as a permutation of $(123...n)$, with a bar over the permutation denoting an helicity flip. For instance, the term of Eq.~(\ref{eq:deltacubeterm}) is invariant under the following 5 symmetries:
\begin{equation}
  345612,\;561234,\;\overline{654321},\;\overline{432165},\;\overline{216543}.
\end{equation}
The former two symmetries are pure permutations, whereas the latter three also involve an helicity flip. These are indeed the symmetries one expects from the three-mass triangle of Figure~\ref{fig:NMHV}.

In the accompanying {\sc Mathematica}~files the results are presented with all symmetries unwrapped to make computations easier. However, to increase readability the symmetries are kept in the formulae in the human readable files, where blocks of spinor helicity expressions are alternated with blocks of symmetries. The convention is that each symmetry in a symmetry block is applied to all the lines in the spinor helicity block preceding it. For example, the NMHV tree amplitude $A_{tree}^{\scriptsize{+-+-+-}}$ of Eq.~(\ref{eq:ApmpmpmOne}) can be written as:
\begin{eqnarray}\label{eq:ApmpmpmOneWithSymmetries}
A_{tree}(1^+,2^-,3^+,4^-,5^+,6^-)&=&
\phantom{+}\frac{1i\langle2|1+3|5]^4}{\langle12\rangle\langle23\rangle[45][56]\langle1|2+3|4]\langle3|1+2|6]s_{123}}\nonumber\\
                                 &&\quad\quad\quad\quad\quad+\;(123456\rightarrow \overline{234561})\nonumber\\
                                 &&\quad\quad\quad\quad\quad+\;(123456\rightarrow 345612).
\end{eqnarray}
Symmetry blocks in general do not contain the full set of symmetries of an expression. Spinor helicity blocks are sometimes symmetric under the missing symmetries, but oftentimes a symmetry is preserved only by the full expression. For instance, the above tree amplitude has 11 symmetries in total, of which only two are used. Among other others, we can find $123456\rightarrow {321654}$, which maps the first spinor helicity line to itself, but also $123456\rightarrow {165432}$ whose action in this case is equivalent to that of $123456\rightarrow \overline{234561}$.

\section{Conclusion}\label{sec:conclusion}

In this article we presented a set of strategies to reconstruct analytical expressions from numerical programs.
We showed how to analyse the singularity structure of amplitude coefficients and parametrise the remaining degrees of freedom in an ansatz. We then described strategies to fit the coefficients in that ansatz. To illustrate our method, we obtained analytical expressions for the six-gluon one-loop amplitude with a gluon in the loop using numerical evaluations from \Blackhat{}. Using these expressions instead of the numerical precedure resulted in a significant speed up. 

The reconstruction strategies presented offer different trade offs between scalability and uniqueness of the result. While the first strategy presented yields a result with a predictable structure, it scales badly with the complexity of the expression. On the contrary, the last two strategies offer more options to control the structure of the outcome, but scale better with the complexity of the problem. An advantage of this flexibility is that it allows one to tailor the form of the reconstructed analytical expression for different goals, such as evaluation speed or numerical stability. For the latter, we can generate equivalent representations of the same expression that are numerically stable in different singular limits.

To conclude, the method presented in this article is not limited to coefficients of one-loop scalar integrals or to numerical algorithms as it can also be used to reformulate existing analytical expressions. For example, it could be used to rewrite analytical two-loop integral coefficients expressed in terms of twistor variables as functions of spinor products, where the pole structure and physical limits are easier to interpret.

\acknowledgments
We would like to thank Adriano Lo Presti for his help in the early stages of this work, and Simon Badger and Ryan Moodie for comments on the draft of this article.

\bibliography{analytical}
\bibliographystyle{JHEP}

\end{document}